\documentclass[fleqn,usenatbib]{mnras}
\usepackage{newtxtext,newtxmath}
\usepackage[T1]{fontenc}

\DeclareRobustCommand{\VAN}[3]{#2}
\let\VANthebibliography\thebibliography
\def\thebibliography{\DeclareRobustCommand{\VAN}[3]{##3}\VANthebibliography}

\usepackage{graphicx}	
\usepackage{amsmath}	
\usepackage{xcolor}
\usepackage{orcidlink}
\usepackage{hyperref}

\title[Light Travel Time Effects in Kilonova Models]{Light Travel Time Effects in Kilonova Models}

\author[F.~McNeill et al.]{F.~McNeill\,\orcidlink{0009-0001-9528-7475}$^{1}$\thanks{E-mail: fmcneill07@qub.ac.uk},
S.~A.~Sim\,\orcidlink{0000-0002-9774-1192}$^{1}$,
C.~E.~Collins\,\orcidlink{0000-0002-0313-7817}$^{2}$,
L.~J.~Shingles\,\orcidlink{0000-0002-5738-1612}$^{3}$,
R.~Damgaard\,\orcidlink{0009-0000-5765-4601}$^{4}$,
A.~Sneppen\,\orcidlink{0000-0002-5460-6126}$^{4}$,
\newauthor
J.~H.~Gillanders\,\orcidlink{0000-0002-8094-6108}$^5$
\\
$^{1}$Astrophysics Research Centre, School of Mathematics \& Physics, Queen's University Belfast, BT7 1NN, Northern Ireland \\
$^{2}$School of Physics, Trinity College Dublin, The University of Dublin, Dublin 2, Ireland \\
$^{3}$GSI Helmholtzzentrum für Schwerionenforschung, Planckstraße 1, 64291 Darmstadt, Germany \\
$^{4}$Cosmic Dawn Center (DAWN), Niels Bohr Institute, University of Copenhagen, Blegdamsvej 17, K{\o}benhavn 2100, Denmark \\
$^5$Astrophysics sub-Department, Department of Physics, University of Oxford, Keble Road, Oxford, OX1 3RH, UK
}

\date{Accepted XXX. Received YYY; in original form ZZZ}

\pubyear{2026}

\begin{document}
\label{firstpage}
\pagerange{\pageref{firstpage}--\pageref{lastpage}}
\maketitle

\begin{abstract}

The extremely rapid evolution of kilonovae results in spectra that change on an hourly basis. These spectra are key to understanding the processes occurring within the event, but this rapid evolution is an unfamiliar domain compared to other explosive transient events, such as supernovae. In particular, the most obvious P Cygni feature in the spectra of AT2017gfo -- commonly attributed to strontium -- possesses an emission component that emerges after, and ultimately outlives, its associated absorption dip. This delay is theorised to arise from reverberation effects, wherein photons emitted earlier in the kilonova’s evolution are scattered before reaching the observer, causing them to be detected at later times. We aim to examine how the finite speed of light -- and therefore the light travel time to an observer -- contributes to the shape and evolution of spectral features in kilonovae. Using a simple model, and tracking the length of the journey photons undertake to an observer, we are able to test the necessity of accounting for this time delay effect when modelling kilonovae. In periods where the photospheric temperature is rapidly evolving, we show spectra synthesised using a time independent approach are visually distinct from those where these time delay effects are accounted for. Therefore, in rapidly evolving events such as kilonovae, time dependence must be taken into account.

\end{abstract}


\section{Introduction}

Observations of the kilonova GW170817/AT2017gfo \citep{abbott2017gw170817, Andreoni2017, Arcavi2017, Chornock2017, coulter_2017, Cowperthwaite2017, Drout2017, Evans2017, Kasliwal2017, Lipunov2017, Nicholl2017, Pian17, Shappee2017, Smartt2017, SoaresSantos2017, Tanvir2017, Troja2017, Utsumi2017, Valenti2017}
show a rapid evolution, with spectral features emerging and changing on a timescale of hours, as recently highlighted by  \cite{sneppen2024emergencehourbyhourrprocessfeatures}. These mergers of two neutron stars are complex events, with the spectral features shaped by the underlying physics of radiative transfer. Understanding how this shapes the features seen in the spectra and their evolution will aid future identification of features and lead to a greater understanding of kilonovae and the processes within. 

\cite{sneppen2024emergencehourbyhourrprocessfeatures} have detailed the rapid evolution of the spectra of AT2017gfo, and argued that light travel time effects can have an important influence on the observed spectrum. Due to the finite speed of light, photons emitted in the same instant are detected by an observer at different times, due to the different path lengths of the photons. Radiation from different parts of the ejecta will arrive at an observer at different times: radiation from the near side will arrive at an observer before radiation from the limb. The scale of such effects is linked to the homologous expansion velocities of the line-forming region, i.e., the effects of light travel times scale with radius, as in homologous expansion $r \propto v$. 

Consider, for example, photons emitted at the same instant (denoted by the time since merger, $t_{\rm ph}$) on a sphere of radius $r$. The detection times of these photons by a distant observer would be spread across a time interval ($\Delta t_{\rm det}$), which will have a maximum duration of 
\begin{equation}
    \label{eqn:v/c}
    \Delta t_{\rm det} = 2 \frac{r}{c} = 2t_{\rm ph} \frac{v}{c},
\end{equation}
since photons emitted from the far side have to travel the diameter of the sphere further than those from the near side. Here we exploit the homologous velocity law to relate the radius of emission ($r$) to the flow velocity ($v = r/t_{\rm ph}$) at the point of emission. We note that Equation~1 gives an upper limit to the spread in detection times since it assumes that photons from the entire emitting surface can reach the observer -- for example, if the emission were from an optically thick photosphere from which the observer sees only the near side, then the spread would be reduced by a factor of two. Nevertheless, it remains true that if emission originates in regions in which $v/c$ is large then the spread in detection times ($\Delta t_{\rm det}$) can be a significant fraction of the time elapsed since merger ($t_{\rm ph}$).

Work by \citet{kasen2006} has shown this effect to be small for supernovae, as ejecta velocities in even the most extreme supernovae are on the scale of $\lesssim 0.1$\,c \citep{finneran2025SNejectav}, hence time-independent radiative transfer codes, such as \textsc{tardis} \citep{KerzendorfSimTardis}, provide good results in modelling these events.
However, the much more rapid expansion velocities of kilonova ejecta ($\lesssim 0.3$\,c, as measured from observations of AT2017gfo; extremes of $\lesssim 0.7 - 0.8$\,c, derived from hydrodynamical simulations of neutron star mergers; \citealt{Smartt2017, watson2019identification, gillanders2022, just2023, kenta2024}) means that light-travel time effects are important for these types of events.

A core technique for identifying features in AT2017gfo spectra has been the fitting of P Cygni profiles to the observations, as in the identification of the \ion{Sr}{ii} \citep{watson2019identification} and \ion{Y}{ii} features \citep{sneppen2023discovery}. Additional plausible identifications include \ion{La}{iii}, \ion{Ce}{iii} \citep{domoto2022} and \ion{Gd}{iii} \citep{rahmouni2025gd}.
P Cygni profiles are comprised of emission centred near the rest wavelength(s) of the line(s) forming them, with absorption blueward of this. They form in outflowing material, and their formation mechanism leads to a strong correlation between the wavelength at which a photon emerges and when they are seen by an observer. Specifically, photons scattered by material expanding towards an observer will encounter a blue-shifted line, and travel shorter distances, meaning they arrive earlier and with bluer wavelengths. However those scattered by ejecta expanding away from an observer or originating from deeper within the event are less blue-shifted. Additionally, those undergoing multiple scattering events have longer path lengths, resulting in later arrival times to the observer. 

Work by \citet{gillanders2024modelling} depicts a clear red-ward drift of observed spectral features as AT2017gfo evolves, which the authors attribute to variations in photon arrival times arising from the rapidly expanding ejecta, with the expansion velocity of the line-forming region decreasing over time. This observed decrease in expansion velocity originates from the ejecta material becoming more dilute as it expands into material ejected with lower velocities, meaning that the photosphere recedes through the ejected material. Furthermore, \cite{sneppen2024emergencehourbyhourrprocessfeatures} highlight the evolution of the \ion{Sr}{ii} feature as being potentially driven by reverberation effects, i.e., when scattered photons take longer paths to an observer, resulting in their arrival being delayed and them becoming red-shifted. In this way, the time delay effect impacts both the temporal evolution and shape of spectral features. Of particular interest is the persistence of the emission component of the \ion{Sr}{ii} P Cygni profile, relative to its absorption dip. We seek to test these ideas by exploring and quantifying light travel time effects in the relativistic velocity regime occupied by kilonovae using a similar Monte Carlo methodology akin to that adopted in e.g., \textsc{tardis}, but with the inclusion of time dependence. We investigate whether the light travel time delay effect can explain the long-lived nature of the observed emission feature in observations of AT2017gfo, and the progression of this feature towards redder wavelengths over time. We note that light travel time effects in kilonovae are already taken into account in time-dependent radiative transfer codes
\citep{kasen2013, kasen2017heavyelements, brethauer2024modellinguncertainties, groenewegen20252dendtoend, tanaka2013, tanaka2017, tanaka2018, wollaeger20212dsupernu, fryer2024velocitydistribution, pognan2022steadystate, pognan2022SUMONLTE, PognanNLTE, bulla2019polarisation, bulla2023possisimprovements, collins2023, shingles2023}.

However, the specific role they play in shaping the spectrum has not yet been discussed in isolation. With this work, we aim to investigate how light-travel time impacts the formation and evolution of spectral features, to guide our understanding of how they appear (see \citealt{pognan2022steadystate}, for discussion on other time dependence effects in kilonova modelling, such as in the radiation field and ionisation rate equations). Since some studies do opt for the time-independent approach \citep[see e.g.,][]{watson2019identification, gillanders2022, perego2022lightelements, vieira2023sparki, vieira2024sparkii, vieira2025sparkiii, mulholland2024sr, Mulholland2024te, mccann2025zr} we seek to understand how impactful this effect is on the conclusions drawn from such works.

\section{Method}

Our approach focusses on modelling the formation of a single, idealised scattering line profile. For concreteness, we choose parameters that are roughly appropriate for the \ion{Sr}{ii} feature identified by \citet{watson2019identification}, but our approach is general and has implications for any early-phase scattering dominated line profile. We model the line-forming region of the kilonova as the volume between two concentric spheres, each expanding as per:
\begin{equation}
    v=r/t_{\rm ph},
\end{equation}
where $v$ is the expansion velocity of the sphere, given its radius $r$, at a time $t_{\rm ph}$ since the merger. 

The inner sphere represents the photosphere -- assumed to emit as a single-temperature blackbody, a choice justified by the work of \citet{sneppen2023blackbody} -- with the outer sphere representing the outer boundary of the ejecta.\footnote{We note that whilst we refer to an ejecta boundary, this outer sphere only needs to represent the extent of the line-forming region, under the assumption that any ejecta beyond this are expected to negligibly contribute.}
We use Monte Carlo Radiative Transfer (MCRT) methods to track the distances travelled by energy packets from their origin at the inner boundary, through any interactions with the spectral line to reach an observer, building on code developed by \citet{noebauer_2019}. An outline of the parameters used in our model follows.

\subsection{Time sequencing} \label{subsec:timeseq}

In the discussion of our results, we will use two clocks and two times. The first, denoted $t_{\rm ph}$, tracks the time when a group of energy packets are initialised at the photosphere, and is defined to be zero at the instant the merger occurs. The second, denoted $t_{\rm det}$, tracks the time since the merger until the packets arrive at the observer, i.e., it tracks the time until the photons are detected. The zero point of this clock is set such that we define the arrival times for photons at the observer relative to the arrival of a (hypothetical) photon released from the centre point of the kilonova at the instant the merger occurred. The times on the two clocks are related through:
\begin{equation}
\label{eqn:time_relation}
t_{\rm det} = t_{\rm ph} + \frac{d}{c},
\end{equation}
where $d$ is defined as the distance to the observer relative to the origin, calculated as per in Appendix A. In this way, $d$ can be negative on the near side of the photosphere.
With this definition, we note that some photons have $t_{\rm det} < t_{\rm ph}$: this occurs because photons emitted from the near side of the photosphere have a negative arrival time correction, since the distance from the origin to the observer is subtracted.

As photons emitted from the photosphere at different times ($t_{\rm ph}$) can reach the observer simultaneously to generate a single spectrum for a given $t_{\rm det}$, a significant $t_{\rm ph}$ interval must be considered. In this study, we consider spectra for a $t_{\rm det}$ range between 1.43 and 4.40 days post-merger, so our calculation consists of a sequence of simulations. This full sequence covers a $t_{\rm ph}$ period from 0.5 to 8 days, with a new group of photons initialised on the receding photosphere every hour of simulation time, which are later combined to create the full spectrum. The selection of our observed $t_{\rm det}$ range -- 1.43 to 4.40 days -- is driven by our choice to assume a blackbody-like photosphere: an assumption only valid and capable of matching the observations of AT2017gfo at these early times \citep{gillanders2022}. This assumption is common in studies utilising codes such as \textsc{tardis}, and is further motivated in kilonovae by the work of \citet{sneppen2023sphericalsymmetry}.

\subsection{Adopted photosphere temperature and velocity evolution} \label{sec:temp+v}
We adopt a simple photospheric model for the radiation entering the line-forming region. Specifically, we assume that radiation enters through the inner boundary with frequency distribution determined by linearly randomly sampling a wavelength range between 2000 and 18000\,\AA\ -- approximately centred around the rest frame wavelength of the \ion{Sr}{ii} multiplet -- and weighting the energy of the packets by a blackbody (of temperature described below) in a reference frame co-moving with the photosphere.\footnote{Included special relativistic corrections are taken to first order (i.e., $v/c$) only; a full treatment is beyond the scope of this work.} The angular distribution is drawn with zero limb darkening of the photosphere, as described by \citet{noebauer_2019}.

We develop a simple profile for the temperature evolution of the photosphere based on the work of \citet{gillanders2022} -- an empirical effort to describe the observations specifically of the event AT2017gfo -- and apply a small shift in temperature. The characteristic temperature of this blackbody is chosen to be appropriate to the conditions present at their release (using methods described below), and this slight shift brings our much simpler model into closer alignment with the observed shape of the AT2017gfo X-shooter spectra \citep{Pian17,Smartt2017} . The temperatures used are detailed in \autoref{tab:fitting_temps}, with any intervening or later temperatures found by interpolating a fourth-order polynomial fitted to these temperatures. This profile was chosen due to providing an adequate match to the observational data following a parameter search.

\begin{table}
    \centering
    \caption{
        Time since the merger occurred and corresponding blackbody temperatures used in the simulation. Temperatures and photospheric expansion velocities for intervening and following times are determined by performing a polynomial fit to the temperatures shown here. The maximum expansion velocity of the material is set to 0.35\,c.
    }
    \begin{tabular}{c|c|c}
        \hline
        $t_{\rm ph}$ (days)     &Blackbody temperature (K)      &Photospheric velocity (c)      \\
        \hline
        0.5 & 10000 & 0.30 \\
        1.4 & 4600 & 0.28 \\
        2.4 & 3600 & 0.20 \\
        3.4 & 3400 & 0.15 \\
        4.4 & 3200 & 0.12 \\
        \hline
    \end{tabular}
    \label{tab:fitting_temps}
\end{table}

To determine the evolution of the expansion velocity of the photosphere, we start from the bolometric luminosities determined by \citet{Smartt2017}. These luminosities are then converted to an approximate corresponding photospheric velocity used at each step of simulation time using:
\begin{equation}
    \label{eqn:Lbol}
    v^2 = \frac{L_{\rm{bol}}}{4 \pi \sigma t_{\rm ph}^2 T^4},
\end{equation}
where $v$ is the photospheric velocity, $T$ is the temperature of the photosphere, $\sigma$ is the Stefan-Boltzmann constant, and $t_{\rm ph}$ the time since explosion. A fourth-order polynomial fit is performed for the intervening times covered by the simulation between those detailed in \autoref{tab:fitting_temps}, matching that used by \citet{gillanders2022}. The maximum expansion velocity of the material was held constant at 0.35\,c throughout.\footnote{Varying this value has a negligible effect on the output of the simulation.}

\subsection{Optical depth} \label{subsec:opacity}

Our modelling focusses on the \ion{Sr}{ii} multiplet which we approximate as a single, pure scattering line -- i.e., there is no fluorescence, and we do not separately account for component transitions in the multiplet. We adopt a single rest wavelength of 10500\,\AA, and we treat line interactions in the Sobolev approximation. Only bound-bound scattering is considered throughout this work following findings by \citet{kasen2013} that other opacity sources provide only minor contributions in kilonovae, and so can be neglected for this exploratory study. We adopt a profile for the Sobolev optical depth that is an inverse power law mimicking the decrease in density with increasing radial distance from the photosphere, described as:
\begin{equation}
    \tau = \alpha \left(\frac{r}{R}\right)^{-\beta},
    \label{eqn:general_power_law}
\end{equation}
where $r$ denotes the radial coordinate and $R$ corresponds to the photospheric radius. $\alpha$ and $\beta$ are free parameters, with $\alpha$ dictating the optical depth at the photosphere, and $\beta$ controlling the rate at which the optical depth falls off with radial distance. $\alpha$ can vary over the course of the simulation. Primarily, we consider a case where the line is always optically thick at the photosphere, and the evolution of $\alpha$ is described by:
\begin{equation}
    \alpha(t_{\rm i}) =
    \begin{cases}
    \dfrac{t_{\rm i}}{1.5} & \text{if } 1.5 < t_{\rm i} \leq 3.0 \; \rm days, \\
    2.0 - 0.5 \times (t_{\rm i} - 3.0) & \text{if } 3.0 < t_{\rm i} \leq 5.0 \; \rm days, \\
    1.0 & \text{otherwise},
    \end{cases}
    \label{eqn:tau_use}
    \end{equation}
where $t_{\rm i}$ indicates the time at which the interaction occurs. These values are chosen to mimic the quick strengthening of the feature at early times followed by the later, slower fading, with the timings determined following a parameter search to match the evolution of the feature in the observational data. This optical depth evolution also captures the recombination of \ion{Sr}{iii} to \ion{Sr}{ii}, allowing it to be followed over the course of the simulation.

Motivated by the work of \citet{sneppen2023recombination}, who argue for a rapid onset of optical depth for the \ion{Sr}{ii} feature, we also explore the effect on the spectra when $\alpha$ is initially set to 0, i.e., no optical depth. At 1.0 days -- when the absorption is first seen by an observer -- the line abruptly ``switches on'', with $\alpha$ immediately raised to 1. However, this did not lead to significant differences to the results or conclusions discussed in this work.

We set $\beta = 3$ throughout,
as it agrees with hydrodynamical simulations of neutron star mergers \citep[][]{Goriely2011, Goriely2013, Goriely2015, Bauswein2013, Hotokezaka2013, tanaka2013}, and has been used successfully in previous works to model the observations of AT2017gfo \citep[e.g.,][]{watson2019identification, gillanders2021constraints, gillanders2022}.

\begin{figure*} 
\centering
    \includegraphics[width = \linewidth]{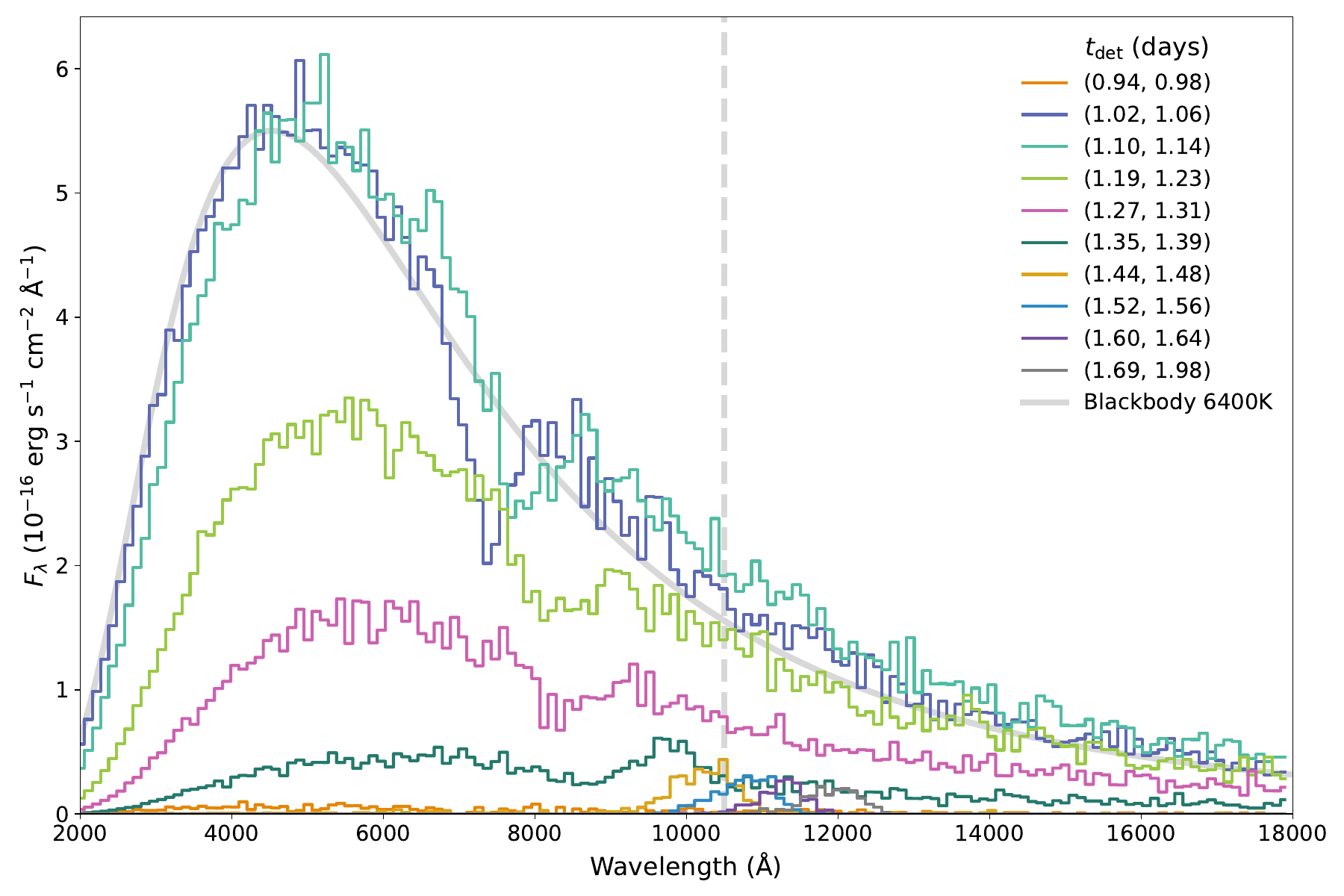}
    \caption{Spectra generated using only packets released from the photosphere at $t_{\rm ph} = 1.43 \; \rm{days}$ post-merger in simulation time. Packets are temporally binned according to the time, $t_{\rm det}$, when they reach a hypothetical observer, with each curve overlaid on each other. The observer is positioned 40\,Mpc from the kilonova, matching the distance to AT2017gfo \citep{Smartt2017}. The time bins are 1\,hr wide. A reference 6400\,K blackbody, chosen to match the peak of the plotted curves, is shown in grey as a comparison aid. For clarity, time bins with low overall counts (with $t_{\rm det}$ between 1.69 and 1.89 days) are grouped together. The rest wavelength of the \ion{Sr}{ii} line is shown with a grey, dashed, vertical line.
    }
    \label{fig:1Dayrelease}
\end{figure*}

\subsection{Line profile calculation}
\label{sec:line_profile}

To simulate the line formation we use a slightly adapted version of the Monte Carlo algorithm from \citet{noebauer_2019}. Using the initial trajectory and frequency of the packets when they leave the photosphere, it is first determined if they will Doppler shift into Sobolev resonance with the line along their flight path through the line-forming region. Photons are limited to only being able to undergo scattering once with a given line due to the assumption of a homologous velocity field: as in such a photon can only ever be redshifted, meaning once it has interacted with the spectral line it will never possess the correct frequency to interact again as here we only consider first-order special relativistic effects (i.e., $v/c$). Additionally, only those photons initialised with a frequency blueward of, and close enough to, the frequency of the line to be able to Doppler shift into it within the ejecta are capable of doing so. If a photon is able to Doppler shift into resonance with the line, the resonance point is determined when the photon reaches this point. The alpha function from \autoref{eqn:tau_use} is then evaluated at this time. Given the line optical depth at that resonance point, we probabilistically determine if a scattering event will occur. If conditions do not allow scattering to occur, the photon continues along its initial path and leaves the model. If scattering does occur, a new direction for the photon is randomly drawn, and it undergoes a frequency change, described by:
\begin{equation}
\label{eqn:new_nu}
    \nu = \nu \frac{1 - \frac{v}{c} \mu_{i}}{1 - \frac{v}{c} \mu_{e}},
\end{equation}
where the frequency of the photon $\nu$ is altered depending on its trajectory incident to the resonance point as defined by the direction cosine, $\mu_i$, and that when it emerges $\mu_e$, in addition to its interaction velocity, $v$. The photon will then continue to leave the model.

If, after scattering, a packet's trajectory leads it to cross the inner boundary, the packet is assumed to be re-absorbed and will not contribute to the spectra. The remaining scattered packets leave the outer boundary and contribute to the spectra. 

This total path length travelled by each photon packet is recorded and converted to a time of flight in the model. In this simulation, a longer path length will correspond not only to a longer delay time, but also a redder wavelength. Photons travelling at an angle and not directly towards an observer will experience a Doppler shift from the expanding photosphere to redder wavelengths. Additionally, a photon scattered from the receding side of the ejecta will be less blue-shifted due to the longer path length they travel from there. 
Furthermore, depending on where the packets emerge on the outer boundary, some will have a greater distance to travel to the observer than others. This relative time of flight to an observer is calculated and added to the distance covered within the boundary to provide an accurate reflection of the evolution of the spectra (see Appendix A). This is combined with the time each photon packet was created to provide a time of arrival at an observer, following which a spectrum can be generated by binning the photons based on the arrival times calculated in this way.

\subsection{Continuous photon flow}
As noted above, a full calculation is composed of multiple simulations, each of which considers a batch of photons ejected at a particular time. The process outlined above will release new photons at regular intervals of one hour in simulation time with appropriate starting conditions for temperature, optical depth, and photosphere expansion velocity. To mimic the effect of continuous evolution when combining these simulations we introduce a small random shift ($\leq\pm30$ minutes of simulation time, drawn randomly assuming a flat distribution) in each photon's time of departure from the photosphere, so that sequential simulations partially overlap in time. This suppresses the effect of discrete timings of release of packets in the calculation to create a more physically realistic continuous outflow. We choose these interval sizes as, following testing, the results were found to be converged.  

\section{Results} \label{sec:results}

\subsection{Photon arrival times}

In this Section, we discuss packets initialised on the photosphere at a fixed time, $t_{\rm ph}$. Here, this $t_{\rm ph}$ is chosen as 1.43 days, which matches the initial X-shooter spectrum of AT2017gfo \citep{Pian17}. \autoref{fig:1Dayrelease} shows all of the packets released from the photosphere at this time in our simulation, binned by the time $t_{\rm det}$, when they are detected. Considering the range of arrival times shown it can immediately be seen that this light travel time effect is impacting the formation of the spectrum. While the familiar shape of the blackbody and beginnings of the formation of a strong feature around 1\,\micron, also seen in AT2017gfo, are present at several observer epochs here, the temporal bins give a strong early indication that light travel time effects are important. When considering the full range of values for $t_{\rm det}$ included in \autoref{fig:1Dayrelease}, we determine that photons released simultaneously from the photosphere have a range of arrival times at an observer spanning 1.25 days, and additionally that photons separated in time bins display substantially different colours. This 1.25 day range arises from the extremes of $t_{\rm det}$ produced by the $t_{\rm ph}$ of 1.43 days; however, we only show a $t_{\rm det}$ range of 0.94 to 1.98 days in \autoref{fig:1Dayrelease} since the flux outside this range comprises $< 1$~per~cent of the total flux. We can calculate the time of flight of the photon packets in this plot using the difference between the $t_{\rm det}$ values and the $t_{\rm ph}$ value of 1.43 days, and compare this to the lifespan of the event. For example: for a packet in the (1.60, 1.64) interval, taking the ratio of this time of flight with the time since the instant of the merger gives a value of 0.13. The time of flight of the photons at this early stage is already over 10~per~cent of the time since the merger. Thus we already illustrate from this initial plot that the time of flight of photons in our simulation is a significant fraction of the lifespan of the event, providing a strong indication that the reverberation effect is non-negligible in the formation of these spectra.

Beyond the large spread of $t_{\rm det}$ depicted in \autoref{fig:1Dayrelease}, we also see rapid spectral evolution on hour-timescales between time bins -- particularly in the those corresponding to 1.04 (purple) and 1.12 days (teal), where the \ion{Sr}{ii} feature is most evident. These curves contribute the most continuum flux at this value of $t_{\rm ph}$. Furthermore, they highlight the incredibly rapid evolution of such events, as discussed in the observed spectra by \citet{sneppen2024emergencehourbyhourrprocessfeatures}. When viewed in comparison to the reference blackbody in \autoref{fig:1Dayrelease}, a clear shift from the deepest absorption and small flux excess of 1.04 days to a sharper absorption and larger flux excess at 1.12 days is present. Additionally, the peak of the continuum flux can be seen to shift to redder wavelengths on this short timescale. This is due to the rapidly changing Doppler shift of the photosphere with time. This change can also be seen in the emission component of the feature: the flux excess above the reference blackbody shifts to redder wavelengths with time. 

\subsection{1.43 days post-merger} \label{sec:1day}

\begin{figure} 
\centering
    \includegraphics[width = \linewidth]{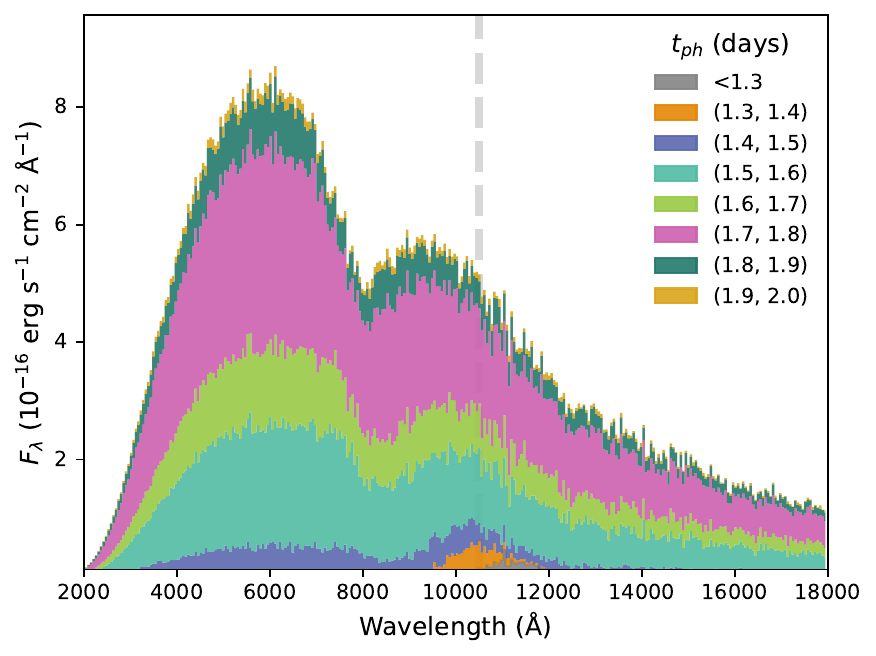}
    \caption{Spectra generated for a three hour $t_{\rm det}$ window centred on the 1.43 day observations of AT2017gfo, located at a distance of 40\,Mpc. The colours correspond to the time of release of packets from the photosphere, $t_{\rm ph}$, with the coloured areas illustrating how many packets released from the photosphere in that time frame are contributing to the spectrum. Each region is stacked to show contributions of the net flux. The grey region consists of time frames with low total numbers of packets, which have been combined and plotted together for clarity. It contains packets in the range $1.0\leq t_{\rm ph} {\rm (d)} <1.3$. The dashed, grey line denotes the rest wavelength of the \ion{Sr}{ii} line.} 
    \label{fig:1Day}
\end{figure}

Having now explained the distribution of $t_{\rm det}$ for photons from one representative $t_{\rm ph}$, we now present full results from our sequence of simulations in which we group all photons by the time at which they reach the hypothetical observer, $t_{\rm det}$ -- located as shown in \autoref{fig:trig} -- in time frames centered around the observational epochs of AT2017gfo from the X-shooter data \citep{Pian17, Smartt2017}. We group packets in 3 hour blocks, i.e., \autoref{fig:1Day} shows the packets from the time frame relating to the 1.43 day spectrum and these packets have arrival times at the observer between 1.3675 and 1.4925 days post-merger. This temporal spacing was chosen as it provides a reasonable compromise between the exposure times of observations from \citet{Pian17} and \cite{Smartt2017}, and the resolution and compute time of the code, as the aim of this work is primarily to provide a simple and computationally efficient model. Convergence testing found a shorter interval does not significantly affect the spectra generated.

The shaded areas of \autoref{fig:1Day} show the contribution to the total flux made by photons initialised on the photosphere at each time in the simulation: a larger area corresponds to a larger flux contribution. The spectrum is cumulative, with the uppermost edge showing the total flux seen by an observer at this time.

This cumulative spectrum shows signs of forming the strong \ion{Sr}{ii} feature identified by \citet{watson2019identification} at this early time: originally seen in the observations of AT2017gfo as a flattening of the blackbody shape in the region of the feature, but this rapidly progresses in a matter of hours to form the absorption dip of the P Cygni profile. In addition to the absorption, we also see the beginnings of the formation of the emission bump, underlining just how quickly the spectra of this event evolve, mimicking the observations of AT2017gfo. At these earliest times, the expansion velocity of the photosphere is highly relativistic -- some studies, notably \citet{sneppen2024emergencehourbyhourrprocessfeatures}, argue for values up to 0.4\,c just a few hours earlier -- smoothing out spectral features. With our model parameters (see \autoref{sec:temp+v}), we may not be capturing the extremes of the evolution of this event.  Whilst we do not fully capture the smooth hour-by-hour evolution seen in the observed spectra of AT2017gfo \citep{sneppen2024emergencehourbyhourrprocessfeatures} due to our choices in time bins, we can still trace the difference a few hours can make in the evolution of the spectra of such an event. The contributions from different $t_{\rm ph}$ in \autoref{fig:1Day} over the course of a few hours are very visually different, e.g.,
if the pink (1.7--1.8\,d time bin) here is compared to the green (1.6--1.7\,d time bin).
Most of the blackbody continuum flux is contributed by photons that were initialised on the near side of the photosphere at $t_{\rm ph} = 1.7-1.8$ days. If the peak of the continuum from these times is compared with that contributed from photons released earlier in the model, e.g., those from $t_{\rm ph} = 1.5-1.6$ days, a difference in the degree of redshifting of this peak can be seen. The packets released at earlier $t_{\rm ph}$ times will have taken a longer path to the observer, and appear towards redder wavelengths since the emitting blackbody is in a frame co-moving with the photosphere, later transformed into the observer's frame. This effect competes with the decrease of the temperature of the photosphere with time, which would lead earlier photons to be emitted from a hotter photosphere, and have bluer wavelengths as discussed by \citet{sneppen2023blackbody}, where the author outlines how different emitting regions of the blackbody will contribute photons which arrive with an observer at different times, and will therefore have different temperatures. However, the central wavelength of the continuum peaks indicate the redshifting effect is slightly stronger at this initial time, with contributions from earlier times in the model centred on fractionally redder wavelengths.

The progression of the shape of the \ion{Sr}{ii} feature can also be traced using the colours in \autoref{fig:1Day} and the progression through time. Photons released with the earliest $t_{\rm ph}$ contribute only around the 10000\,\AA\ range, as these are predominately those packets which have undergone scattering in our model, leading to a longer path length to the observer. In this way, as all photons comprising this spectrum have similar $t_{\rm det}$ values, the smallest $t_{\rm ph}$ values here will have the biggest difference between $t_{\rm det}$ and $t_{\rm ph}$, i.e., they have the longest travel times to the observer. Although several earlier $t_{\rm ph}$ bins have been grouped and plotted together for clarity (in grey), they can still be seen to predominantly contribute to the spectrum around the region of the feature, due to the scattering. It is notable that there is such a wide spread in the time frames contributing to this spectrum -- not only for the formation of the \ion{Sr}{ii} feature but also for the continuum flux -- clearly indicating that at early times, time independent codes may not offer the full picture of spectral evolution. Furthermore, when we consider the time range for which scattered photons dominate a time segment of \autoref{fig:1Day}, and compare that to when the continuum contribution dominates, we see a difference of only 0.1 days: i.e., the green segment of this plot is scattering dominated, but even just a few hours later, the pink segment contains a large amount of continuum photons. This implies a small line scattering region at this time, with most interactions occurring close to the photosphere, and these scattered photons not being significantly delayed compared to the unscattered ones.

\begin{figure} 
\centering
    \includegraphics[width = \linewidth]{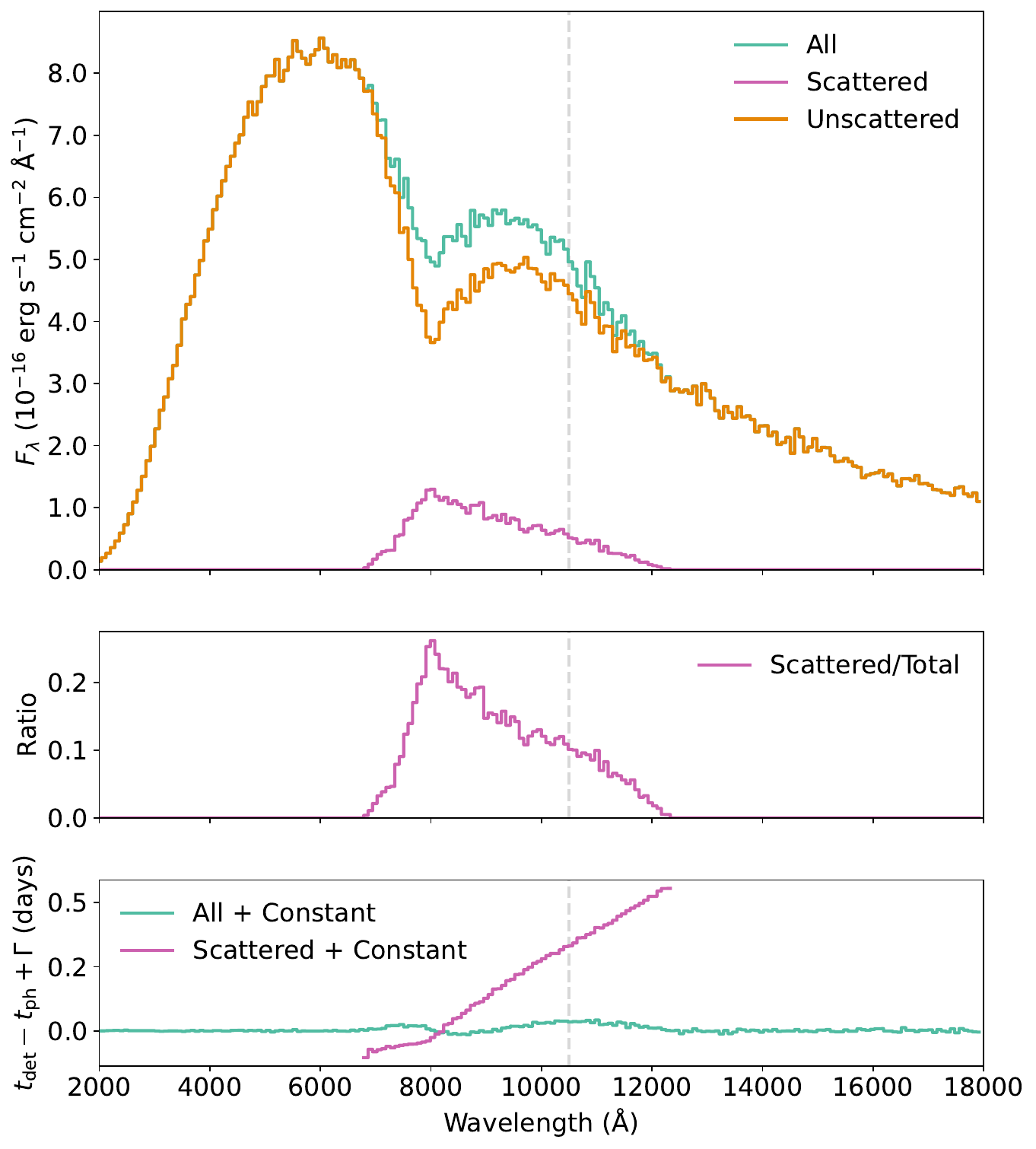}
    \caption{Spectra generated under conditions chosen to match the 1.43 day observations of AT2017gfo. Top panel shows the total flux emitted in the 3 hour window centred on the 1.43 day observation (teal), which is then subdivided into separate curves showing the scattered (pink) and unscattered (orange) photons. Middle panel shows the ratio of the photon flux that was scattered to the total number of photons in each wavelength bin. The bottom panel shows the average difference between $t_{\rm ph}$ and $t_{\rm det}$ of the scattered photons (pink) in each wavelength bin. This is shown relative to this difference calculated for all photons (teal) with a constant, $\Gamma$, added to bring this baseline to 0 on the y axis. $\Gamma$ is determined by calculating the average delay experienced by the unscattered photon packets in the $t_{\rm det}$ range plotted in this figure. Here, $\Gamma$ has a value of 0.28 days. The dashed, grey line denotes the rest wavelength of the \ion{Sr}{ii} line.}
    \label{fig:1Day3panel}
\end{figure}

\autoref{fig:1Day3panel} illustrates a more detailed breakdown of the scattered packets by wavelength bin. From the top panel, it is clear that the packets scattered in the simulation are concentrated around the \ion{Sr}{ii} feature, as is expected. However, the peak of the curve illustrating only the scattered photons does not align with the centre of the emission bump. The large excess of photons in the peak of the blackbody compared to that of the tail, combined with the code not differentiating between those photons scattered to redder wavelengths, and those forward scattered to bluer wavelengths, is the cause of this peak being bluer than the centre of the emission bump. This panel therefore highlights that even in the blue-shifted region of the P Cygni feature, many of the photons here were scattered.

Although the feature does not yet have the strong shape it will later possess, the middle panel in \autoref{fig:1Day3panel} illustrates the contribution that scattering effects are already making to the spectrum, where up to 20~per~cent of photons are scattered near the absorption dip. \citet{sneppen2024emergencehourbyhourrprocessfeatures} finds the observed 1\,\micron\ absorption feature emerges at a time of 1.17 days post-merger, and this work points towards a similar conclusion: that the interactions causing this feature are visible from very early times due to the large fraction of scattered photons.

The third panel of \autoref{fig:1Day3panel} begins to paint a picture of the relevance of the reverberation effect on the formation of the \ion{Sr}{ii} feature. This plot is constructed to illustrate the average time delay between a photon packet leaving the photosphere, $t_{\rm ph}$, and when it then arrives at the observer, $t_{\rm det}$ for each wavelength bin. The baseline value of this delay for all photons (scattered and unscattered) is defined to give an average value of 0. However, we do see an obvious drop below this baseline value for some scattered photons at bluer wavelengths. This is a physical consequence of the homologous expansion of our photosphere. At the earliest times when the photosphere is still small, for a photon scattered near the photosphere and into the line of sight of an observer, it is likely to arrive at an observer before an unscattered photon emitted from the limb. This effect is most prominent at early times, as there are a wider range of emergent trajectories a scattered photon can have to arrive before an unscattered photon while the photosphere is small. As the photosphere expands, this effect will decrease (see \autoref{subsec:later_times}). Already, for the 1.43 day spectrum, it can be seen that the photons scattered to the longest wavelengths experience a time of flight of as much as half a day longer than those continuum packets at similar wavelengths. Even at this early time in the simulation, photons are already being observed that had their origins at significantly earlier times, a solid indication that this reverberation effect should not be discounted in modelling such events.

\subsection{Later times: 2.42 - 4.40 days} \label{subsec:later_times}

Here we aim to track the reverberation effect through the evolution of the kilonova, thus we now look to later times to track this progression. In this Section we will consider two epochs: those centred on 2.42 and 4.40 days post-merger. We select these epochs to show the rapid progression across a day from 1.43 to 2.42 days and focus on 4.40 days as this is the latest time where we consider the photospheric approximation to be valid (see \autoref{subsec:timeseq}). As we wish to understand the importance of the reverberation effect, it follows that we pay particular attention to the latest time-frame possible since this gives the best case to consider the persistence of the emission. Additionally, the overall decreasing flux at later times may lead to "late" arriving photons from earlier times being a more substantial fraction of the photons being detected, creating a more prominent red wing of the emission feature, and we seek to understand this impact. In this way, the trends seen due to the reverberation effect in models of AT2017gfo may be used to predict how features may appear in the spectra of future events. At early times, when the kilonova is bright, a larger number of photons are present to be scattered and these photons are detected in higher fractions later. Hence this effect of redshifting of features due to the reverberation effect will be increased in fainter, faster kilonovae than those which evolve more slowly. We exclude discussion of the 3.41 day epoch for brevity and because the relevant estimates of time delay effects are bracketed by the earlier and later epochs.

\begin{figure} 
\centering
    \includegraphics[width = \linewidth]{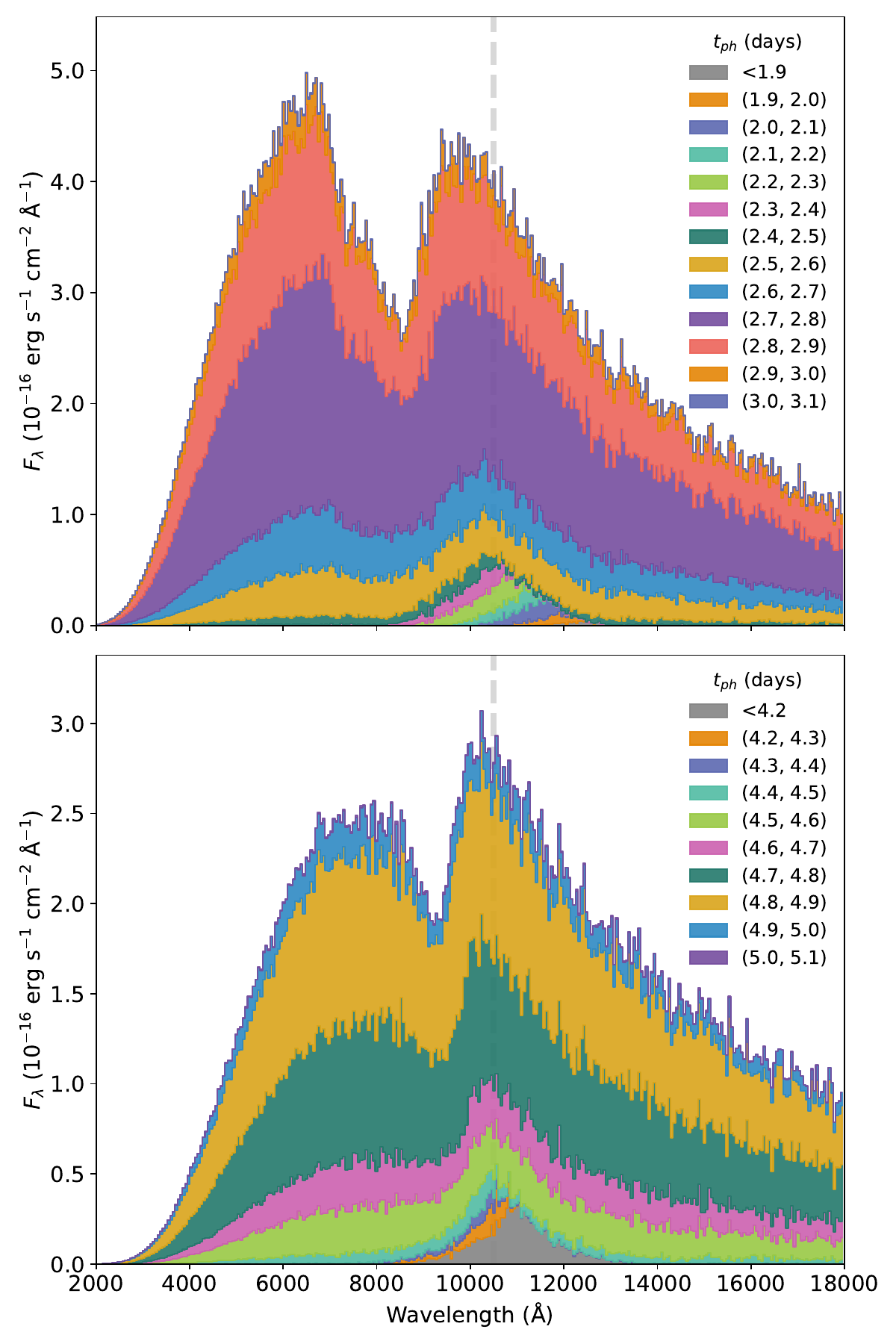}
    \caption{As for \autoref{fig:1Day}, but instead centred around the 2.42 day observational time frame (upper panel) and the 4.40 day observational time frame (lower panel). In the upper panel, the grey region denoted $< 1.9$ in the legend contains packets within the range $1.6\leq t_{\rm ph} {\rm (d)} < 1.9$, and for the lower panel, the equivalent range is $2.8\leq t_{\rm ph} {\rm (d)} < 4.2$.}
    \label{fig:2Day}
\end{figure}

\begin{figure*} 
\centering
    \includegraphics[width = \linewidth]{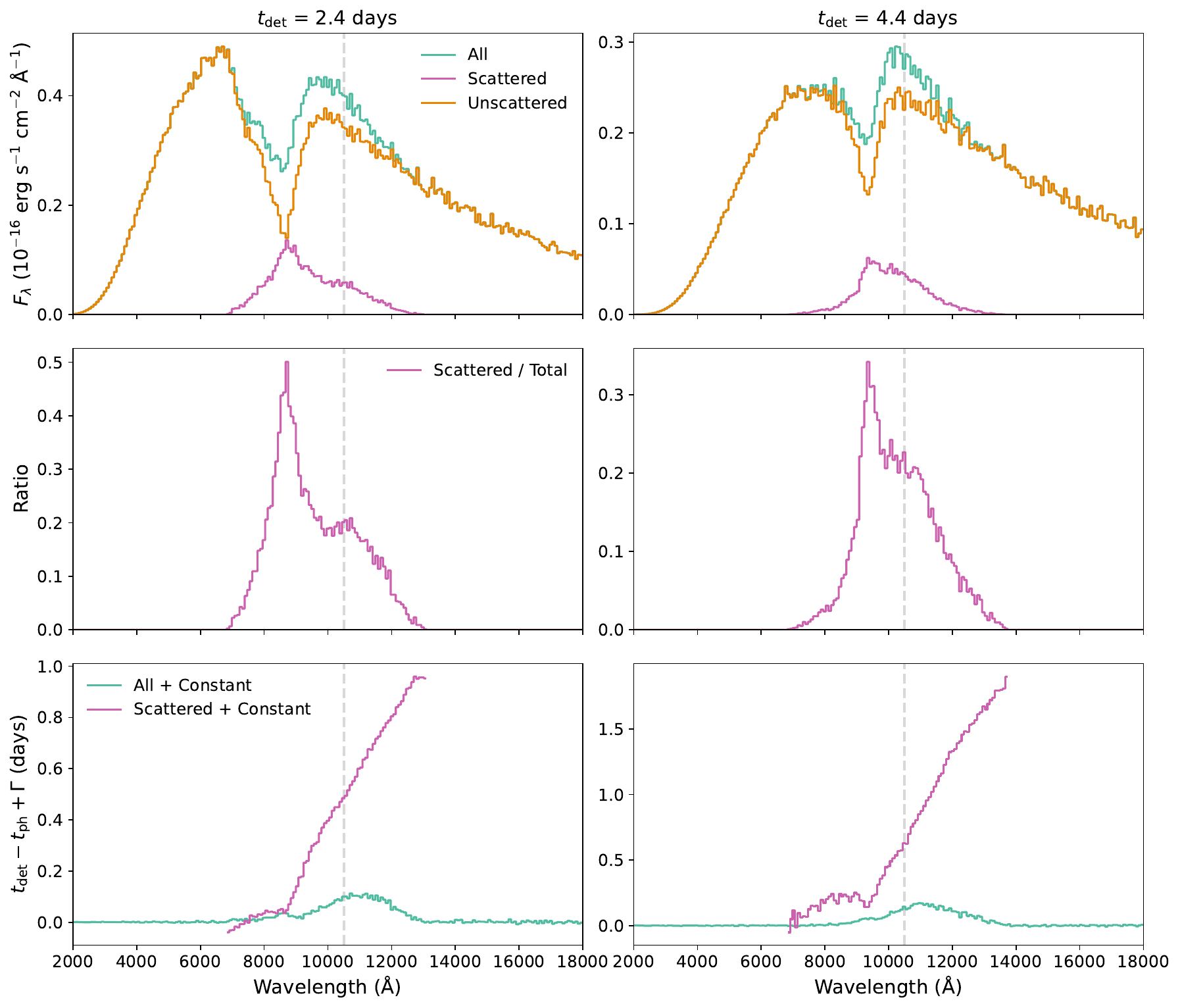}
    \caption{As for \autoref{fig:1Day3panel}, with the left column centred around the 2.42 day observational time frame, and the right column centred around 4.40 days. Values of $\Gamma$ used in the bottom row are 0.35 and 0.36 days respectively.}
    \label{fig:3panel2+4days}
\end{figure*}

For our discussion we present plots similar to those in \autoref{sec:1day}: \autoref{fig:2Day} shows the corresponding $t_{\rm ph}$ for photons arriving in a 3 hour $t_{\rm det}$ window centred on 2.42 and 4.40 days. The two columns of \autoref{fig:3panel2+4days} break down the contributions of scattered photons to the overall spectrum in the same $t_{\rm det}$ time frames. As time progresses since the moment of the merger, the absorption dip progressively deepens, and in the observer's 2.42 day time frame a strong, redder, emission component has formed. Considering \autoref{fig:2Day}, we can clearly see the packets with longer times of flight, i.e., those with earlier release times, arriving and strengthening the emission at this time. An initial hint that the emission spike will be very long lasting when compared to the absorption dip in this feature is the range of $t_{\rm ph}$ contributing to the emission. In \autoref{fig:2Day}, we illustrate an approximately 0.5 day range (2.0 < $t_{\rm ph}$ < 2.5 days) contributing to only emission in the line, dominated by scattered photons, which is reinforced with the later, synthetic spectra. However, referring back to \autoref{fig:1Day}, we see a similar contribution for a $t_{\rm ph}$ range of only 0.1 days.

We additionally see a continuation of the redshifting of scattered photons established in \autoref{fig:1Day} earlier. The photons released from the photosphere at the earliest times shown are located in the \ion{Sr}{ii} feature as within the constraints of this model, a direct continuum photon cannot be delayed this much. Therefore these photons must have been scattered close to the line wavelength. These photons are also affected by their longer travel times being caused by longer path lengths which results in a higher degree of redshift. We see this progressive redshift very clearly here in both panels: if the grey, orange, purple, and teal regions representing the smallest $t_{\rm ph}$ values are considered, the earliest photons released are very clearly red-ward of those released after. Looking at the peak of the blackbody continuum flux we see this effect in competition with the temperature evolution of the kilonova. Photons released earlier result in a flux contribution that is redder in wavelength due to the travel distance and time to the observer than those released later. However, the temperature evolution -- wherein the packets released earlier from a hotter photosphere will be bluer -- can be seen to compete with this redshifting due to path length, resulting in minimal shift to the blackbody continuum peak of both panels in \autoref{fig:2Day} for the range of $t_{\rm ph}$ presented. 

We now focus on the bottom panel of \autoref{fig:2Day}, constructed in the same manner, but centred on a $t_{\rm det}$ of 4.40 days to illustrate the short-lived nature of the absorption dip of the feature. In comparison to the top panel, we see the absorption component of the P Cygni feature has become narrower and shallower while the emission spike continues to strengthen, a clear sign that the spike will persist long after the absorption is no longer visible. If the shape of both the absorption and emission components of the feature are compared across both panels, we see both becoming sharper. This is a consequence of the evolution of the optical depth throughout our model: as $\tau$ decreases when $t_{\rm ph}$ reaches beyond 3 days, lower velocities are present, affecting the shapes of both the absorption and emission profiles here. This is most clearly seen in the emission: at a $t_{\rm det}$ of 2.42 days the top panel shows a smoother bump, which by 4.40 days in the bottom panel has sharpened and become a more localised peak.

We additionally look to the full range of $t_{\rm ph}$ contributing to forming \autoref{fig:2Day}. As in \autoref{fig:1Day}, for clarity several $t_{\rm ph}$ lower signal time frames have been grouped together and plotted in grey. These have been separated into those with a $t_{\rm ph}$ less than the earliest high signal time bin, and those with a $t_{\rm ph}$ greater than the latest high signal time bin. Throughout \autoref{fig:1Day} and \autoref{fig:2Day}, these $t_{\rm ph}$ ranges can be analysed by calculating the ratio between the spread of $t_{\rm ph}$ contributing to each plot, $\Delta t_{\rm ph}$, to the observer time frame, $t_{\rm det}$. As time progresses through the simulation, the $\Delta t_{\rm ph}$ spread for each plot increases, but so too does the $t_{\rm det}$. Comparing this $\Delta t_{\rm ph} / t_{\rm det}$ ratio for each plot provides insight into how the role of significantly delayed photons changes at each time. 
Considering the bottom panel of \autoref{fig:2Day}, this plot contains flux from as early as $t_{\rm ph} = 2.9-3.0 \; \textrm{days}$ contributing to the $t_{\rm det}$ of 4.40 days. This is already quite a significant portion of the time since the merger occurred, although it is a less dramatic effect here than at earlier times. Comparing this plot to the earlier \autoref{fig:1Day} makes this effect even more obvious: both plots have a range of $t_{\rm ph}$ contributing to the spectra of approximately 0.7 days, however this is a more significant fraction of 1.43 days than of 4.40 days. With the decrease in photospheric velocity with time comes a decrease in the fraction of time since explosion where photons will contribute to the spectrum. The delayed photons are clustered at the location of the \ion{Sr}{ii} line, with the most delayed shown at this later $t_{\rm det}$ being contained there due to the geometry in this model, and the feature is shaped by a combination of the evolution of luminosity and velocity. The lower velocities in the model at later times produce sharper features, and the decrease in emitted flux emphasises the contributions from earlier times. This evolution underlines the importance of the reverberation effect in understanding the formation and evolution of spectral features in kilonovae. 

Considering \autoref{fig:3panel2+4days}, we again see the trends established in \autoref{fig:1Day3panel} persisting and becoming more exaggerated. For both $t_{\rm det}$ of 2.42 and 4.40 days -- plotted now in the left and right columns, respectively -- there is now a more defined and obvious difference between the full spectrum and the curve showing the contribution of the unscattered photons in the top panels. It can also be seen from the top panels that the peak of the scattered photons has already shifted to redder wavelengths, although still remains bluer than the rest wavelength of the line. This blue-shifting is an occultation effect: with all of the approaching hemisphere visible to an observer, versus some of the receding hemisphere hidden behind the photosphere, blue shifted emission is preferentially seen. For this feature, and considering initially a $t_{\rm det}$ of 2.42 days, the middle left panel of \autoref{fig:3panel2+4days} illustrates how the ratio of scattered to unscattered photons reaching an observer has approximately doubled for a $t_{\rm det}$ of 2.42 days. Looking to the bottom left plot of \autoref{fig:3panel2+4days}, the average time delay of a scattered photon, has also approximately doubled compared to the same plot for the 1.43 day $t_{\rm det}$ frame, with the reddest scattered photons already drifting to longer wavelengths. Although a bigger redshift at earlier times would be expected due to higher velocities, in these early stages of our model the optical depth of the \ion{Sr}{ii} line is increasing. This results in more scattering, in combination with the photons travelling large distances post scattering from earlier times beginning to arrive at the observer, and these photons will naturally be redder than the early ones. 

Focussing now on the middle right panel: whilst the ratio of scattered photons of the total present in each wavelength bin has decreased at the peak of this plot, the long red tail of this curve illustrates the broader range of wavelengths that photons are scattered to at this time. The cut-off seen for the reddest wavelengths here is related to the signal-to-noise ratio of the simulation. The optical depth at velocities corresponding to the maximum included in the simulation is not zero, however it is very small. Therefore, the probability of photons scattering is also low. A bigger contribution from photons that travel longer distances is possible with time, but occurs in low quantities in our simulation.

\begin{figure*} 
\centering
    \includegraphics[width = \linewidth]{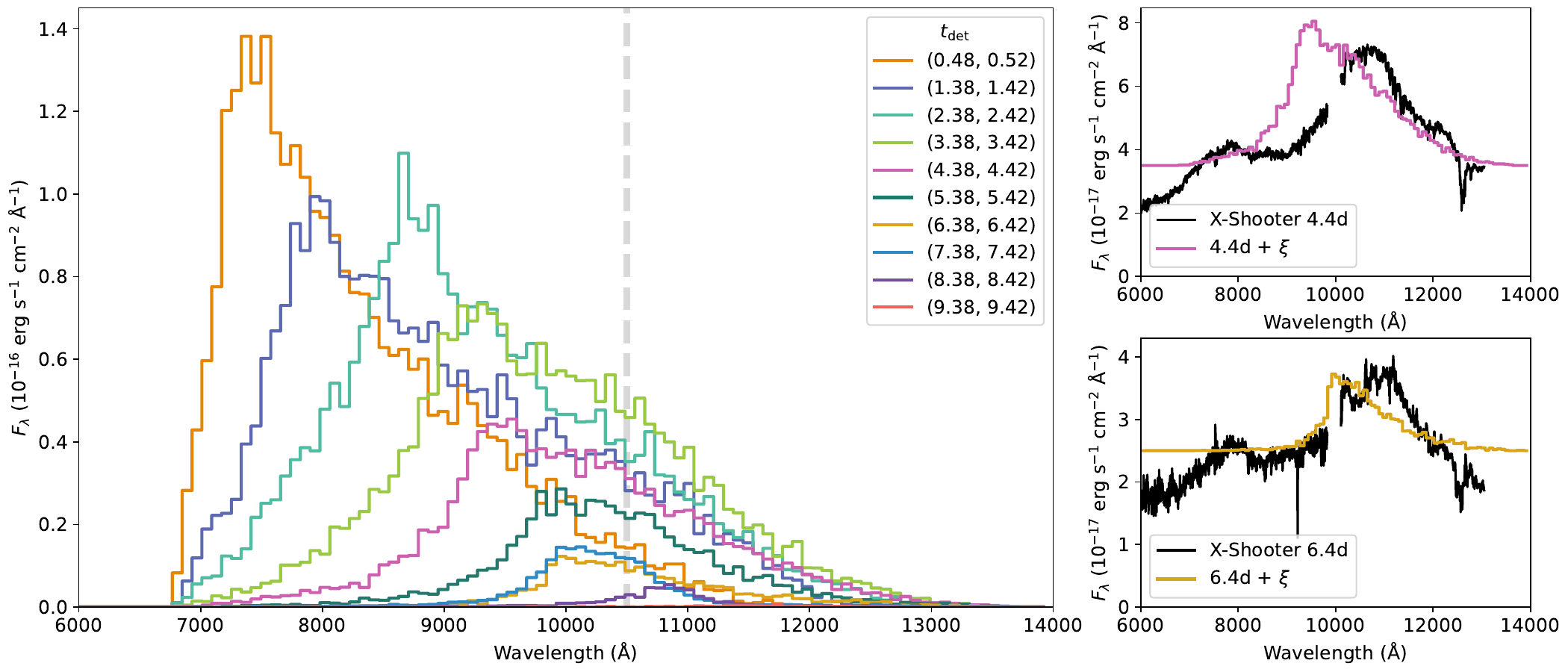}
    \caption{This plot isolates only the scattered packets arriving at the observer in a 3 hour window centred on the observational epochs, illustrated by the changing curve colours. A clear red-ward drift of the peak flux can be seen as time progresses. The dramatic decrease in peak flux as time progresses is attributable to the simulation ceasing to release new packets from the photosphere at 8 days post-merger in the co-moving photospheric frame. No further photons are injected after this time. From this point, the only photons reaching the observer in later time bins are those with long enough scattering distances to increase their time of flight sufficiently. $\xi$ is an arbitrary constant added in place of a full continuum for comparative purposes. The dashed, grey line in the main panel shows the rest wavelength of the \ion{Sr}{ii} line.}
    \label{fig:scatteronly}
\end{figure*}

One of the more striking results from this analysis can be seen in the bottom right panel of \autoref{fig:3panel2+4days}. As before, this plot illustrates the average time delay of the scattered photons versus those forming the continuum. The reverberation effect is visible here: at a $t_{\rm det}$ of 4.40 days post-merger, the observer is seeing photons originating from the photosphere more than 1.5 days previously (included in the <4.2 day contribution in the bottom panel of \autoref{fig:2Day}, containing a broad spread of $t_{\rm ph}$) meaning these packets would have been travelling for around 35~per~cent of the time since the merger occurred to reach the observer, and have been significantly red-shifted. The packets at this extreme now possess wavelengths around 1000\,\AA\ longer than in the comparative plot to the left. This can, however, be attributed to probabilities within the model: photons are not prohibited from being detected with redshift corresponding up to $v_{\rm max} / c$, however this occurs in small quantities and is therefore not visible in the outputs. This strong scattering is also causing a visible change in the teal curve here, depicting all of the packets, both scattered and unscattered. The scattering -- and therefore reverberation effect -- here is strong enough to cause a visible bump in the normalised spectrum at around 11500\,\AA\ at roughly double the height seen to the left at a $t_{\rm det}$ of 2.42 days, a baseline constructed to be roughly 0 on this scale. 

\subsection{Scattering trends}

As a major aim of this paper is to understand the extent to which the persistent red emission and redward drift of the emission features in the spectra of AT2017gfo -- as highlighted by \citet{gillanders2024modelling} -- can be explained by the reverberation effect, we here focus on just those packets in our model which are scattered on their journey to the observer.

To further illustrate the strength of the scattering shown in this simulation, \autoref{fig:scatteronly} isolates only the packets that undergo scattering throughout the simulation, with the different curves illustrating when these packets arrive at the observer. We again use each curve to represent photons arriving within a time frame of 3 hours centred around each observational epoch for AT2017gfo in the observer's frame of reference. We additionally provide a comparison to the observational X-shooter data for two epochs to determine if time delays from scattering alone can explain the shape and persistence of the feature. 

We find a move towards redder wavelengths for the peak of the feature throughout the epochs shown in \autoref{fig:scatteronly}. The flux contribution from scattered photons decreases as time progresses, due to the rapid cooling and decrease in overall flux emitted from the kilonova. This underlines our earlier discussion in \autoref{subsec:later_times}. At the early, bright times in the kilonova's evolution, a large number of photons are scattered, producing a strong peak in \autoref{fig:scatteronly}. As the kilonova fades, a smaller number of photons are scattered and the fraction of scattered, redshifted photons detected from earlier times is higher (see evolution between columns in \autoref{fig:3panel2+4days}). This move towards detecting a higher fraction of redder, scattered photons may partially explain the red-ward evolution of features in the observed spectra of AT2017gfo, discussed by \citet{gillanders2024modelling}. However, when our model is compared to the observed spectra (as plotted to the right of \autoref{fig:scatteronly}) we find the peak flux is too blue to fully replicate the feature in the data. It is additionally too bright at 4.40 days. Since the scattered photons in our model provide a better match to the observed data at later times, this bluer scattering can be attributed to the simplicity of our model and the occultation effect as described in \autoref{subsec:later_times}. Additionally, we find the reduction in overall flux with time is too extreme to fully explain the persistent strength of the observed feature: there are simply not enough scattered photons from the reverberation effect alone. Whilst we find the scattered flux is not negligible at these times, it is not strong enough to be the sole explanation for the evolution of the feature seen in the data.

\begin{figure*} 
\centering
    \includegraphics[width = \linewidth]{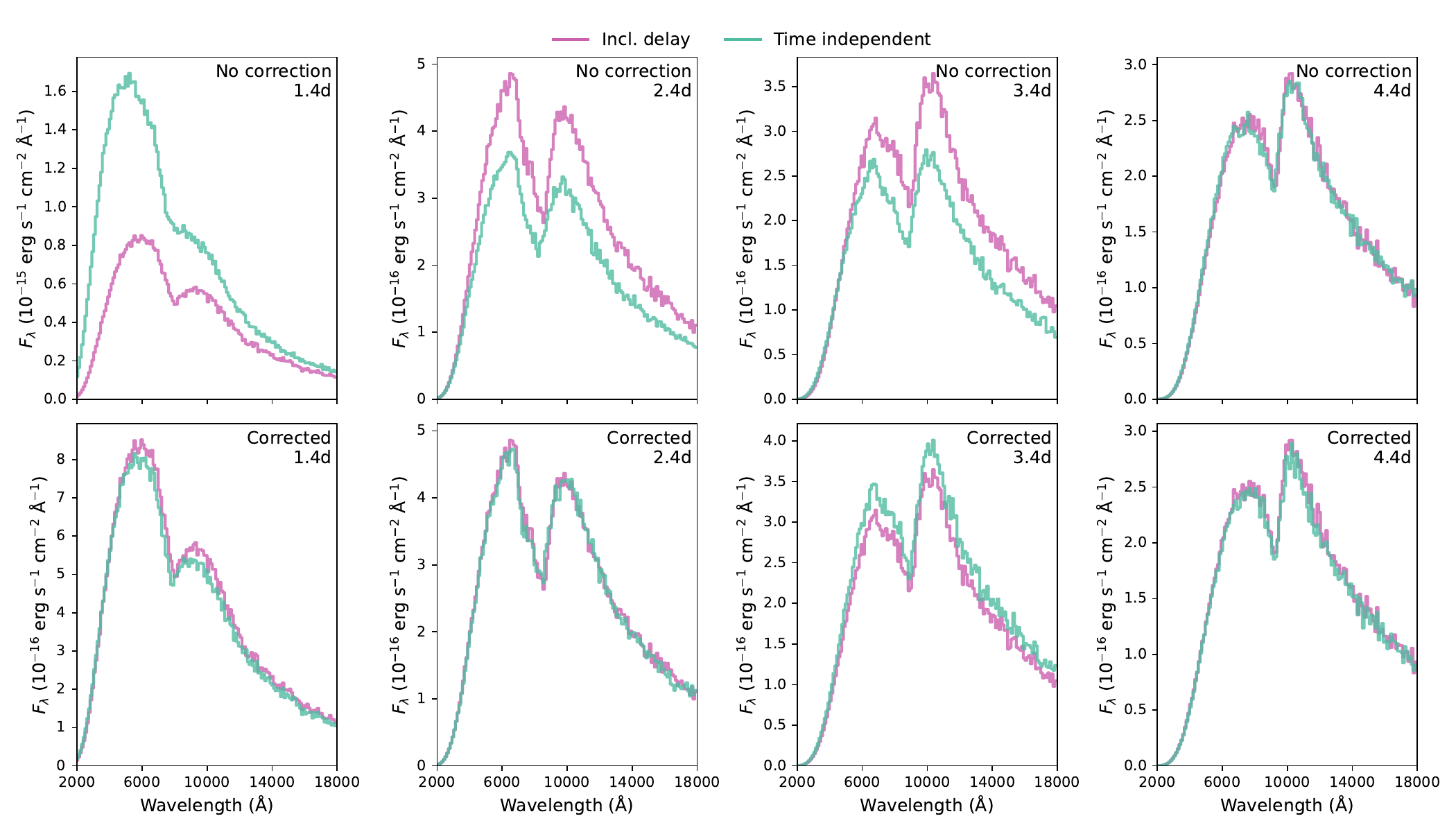}
    \caption{Here we illustrate the effect photon scattering can have on the photospheric temperature inferred by an observer, for four 3 hour windows, centred on the observational epochs from 1.43 to 4.40 days. The top row provides a comparison between this work, accounting for light travel time effects (pink) and a simpler model assuming the speed of light to be infinite (blue). The bottom row indicates similar, but here we add the average delay experienced by energy packets at each observational epoch from our model to the results of the simpler model (blue) to roughly approximate this light travel time effect. From left to right, the values of these corrections are -0.27, -0.34, -0.35, and -0.35 days. We again compare to the work presented in this paper (pink).}
    \label{fig:snap_v_delay}
\end{figure*}

Furthermore, at later times, the red tail of the scattered photons in \autoref{fig:scatteronly} is shifted to redder wavelengths than illustrated by \citet{gillanders2024modelling}, though we do agree with their findings that the feature fades away at 7.4 days post-merger. Beyond this, we find the flux to be too small and too redshifted to contribute meaningfully to the spectrum. As such, we conclude that while the reverberation effect certainly contributes to the evolution of spectral features -- and this effect may be more pronounced in more rapidly evolving, fainter events -- it cannot fully explain the persistent strength of the 1\,\micron\ feature observed in AT2017gfo, though it does warrant consideration in future modelling efforts.

\subsection{Simple time delay correction}

We illustrate the difference accounting for this light travel time makes versus using a time-independent modelling code in \autoref{fig:snap_v_delay}. Here we show a series of spectra from 1.43 to 4.40 days post-merger again, within the 3 hour $t_{\rm det}$ windows used throughout. Within \autoref{fig:snap_v_delay}, the pink curve shows the results of the modelling code used throughout this work, which accounts for the travel time of the photons to an observer. The teal curve, however, ignores this effect and assumes a photon packet released from the photosphere arrives instantaneously at an observer, mimicking the result that might be obtained from a time-independent approach. The top row of spectra provides a straightforward comparison between these two approaches: both are simply plotted together on the same axis for the relevant epoch. If the light travel time effects discussed throughout this work had little or no impact on the spectra produced, we would expect both the pink and teal curves plotted in this top row of \autoref{fig:snap_v_delay} to be visually very similar. However, it can be seen that -- while the shapes of the spectra broadly agree across the time depicted -- there are significant differences in the level of flux predicted by the two approaches at some epochs. 

There is a large discrepancy in the overall flux level seen at 1.43 days, with the model accounting for time delay effects having a peak flux approximately 50~per~cent of that from the time-independent model. Additionally, at 3.41 days post-merger, we see a difference between the height ratio of the peak of the continuum to the peak of the feature arise between the two models. These large differences are due to the rapid temperature evolution of the photosphere: when the temperatures here change on timescales comparable to the time of flight of a photon packet to an observer, the time-independent approach to modelling is less accurate. Due to this rapid change, in combination with light travel times, an observer is not seeing spectral contributions from a single moment in time, but rather a range of times and temperatures. Conversely, the 2.42 and 4.40 day spectra presented here do indeed look very alike, and this is likely due to the fairly simplistic way we treat the temperature evolution of the blackbody throughout this work. Here, the temperature is evolving more slowly, so an observer sees contributions to the spectrum from a smaller range in $t_{\rm ph}$ times, improving the accuracy of the time independent approach. Therefore, we can conclude that the difference due to light travel time delays is largest when the temperature is evolving most rapidly, i.e., at 1.43 days in this case. 

We also note a reversal in which model produces the brightest spectra over time. Initially, at 1.43 days, the time independent model produces a brighter spectrum, with a switch to the time dependent spectrum being brighter at 2.42 days and beyond. The luminosity of the kilonova is linked to both the temperature of the photosphere raised to the fourth power and the emitting radius squared. Since both of these variables evolve with time, the model that appears brighter is due to whichever of these two variables dominated the evolution at that point. The geometry of our model is such that the delay times for a photon travelling to an observer are always negative, i.e., for a photon released from the nearside of the photosphere, $t_{\rm det} - t_{\rm ph}$ will always be  negative, meaning the time dependent model is always seen later in photosphere time. At 1.43 days, the rate of decrease of the temperature is dominant over the rate of expansion of the photosphere resulting in a time independent spectrum which is much brighter, since this is a spectrum from an earlier, hotter, photosphere. At later times, both the rate of change of temperature and rate of change in photospheric radius becomes much less dramatic. At these times, a pure blackbody follows the established pattern: the time independent model, seen at earlier photosphere times is brighter. However, complications arise from the inclusion of line interactions. This complexity leads to the time independent model appearing dimmer than the time dependent model. These differences indicate that whilst time-independent modelling codes can still be used to produce spectra that match observations, some consideration should be given to the temperature profile of such spectral evolution, particularly during periods of rapid temperature evolution.

To further illustrate this point, in the bottom row of spectra shown in \autoref{fig:snap_v_delay}, we apply a very simple time correction to the time independent results and show that this correction brings the spectra into good agreement. This correction is performed by calculating the average travel time ({$t_{\rm det}$ - $t_{\rm ph}$)} of a photon packet for each epoch in our model contained within the 3 hour $t_{\rm det}$ windows plotted here, and then adding that average travel time to each photon packet in the time-independent model. This technique effectively moves the selection window for the photons in the time-independent model: e.g., some photons with a $t_{\rm det}$ falling between the 3 hour windows centred on 1.43 and 2.42 days will then fall into the 2.42 day window with this correction applied. Whilst a crude technique, the bottom row of spectra in \autoref{fig:snap_v_delay} are much more similar than the top row. 

The spectra are still not completely in agreement, as can be seen at 1.43 days where the temperature is evolving rapidly and 3.41 days where scattering is beginning to impact the spectral shape, but this simple correction moves them closer to that point. Focussing on this 1.43 day spectrum, it follows that the rapid evolution of photospheric temperature leads to an observer seeing photons emitted from a range of different photospheric conditions in quick succession. This makes it a difficult task to decide on one simple temperature shift which provides an adequate correction to the synthetic spectrum. Therefore, at these times of rapid photospheric evolution, a more complex treatment would be necessary to increase this alignment, but for our illustrative purposes this simple correction is sufficient. Adding this delay effectively shifts the point in the temperature evolution the observer sees. Without this correction, the observer sees a photosphere much brighter and bluer using the time-independent modelling method, as can be most clearly seen in the top left plot of \autoref{fig:snap_v_delay}. Comparing this to the bottom left plot, post-correction at the same epoch, it can be easily seen that the continuum flux produced is much more similar between the two methods. 

Interestingly, it emerges that this correction is more effective in the redder wavelengths for this model. This is due to a range of temperatures contributing to the continuum, and so this wider spread of temperatures makes the correction less good for bluer wavelengths here. If the range of $t_{\rm ph}$ values contributing meaningfully to the continuum through this simulation is considered, we find a strong contribution across a period of 0.4 days. Across this time period so early in the evolution of this event, the temperature will be changing extremely rapidly, so such a simple correction as used in this model cannot capture the nuances of this time period. This is reinforced at later epochs, as the spectra show better alignment between the two modelling methods across the full wavelength range, without the separation between blue and red, which we can conclude is due to the slower temperature evolution here, making the simple correction more effective. Additionally, \citet{sadeh2025} discusses the need for relativistic effects to be taken into account when modelling these events as failing to do so artificially inflates the ejecta temperatures and expansion velocities inferred from fitting attempts, reinforcing the care that should be taken to not neglect these effects. While this does not invalidate work done using these time-independent codes, we stress that this effect warrants consideration when drawing conclusions from such models. 
The effect of such a simple correction is striking. As illustrated by the plots for the 1.43 day spectra in \autoref{fig:snap_v_delay}, in some cases the observed spectrum does not accurately reflect the emitted spectrum at one specific time since due to this rapid evolution an observer sees a meaningful contribution from a range of different photospheric times and temperatures. Therefore, care should be taken to consider the time delay effect involved in light travel, particularly when drawing conclusions related to the temperature of the continua of kilonovae in phases where they evolve most rapidly, as the impact of this effect is strongest here. 

\section{Conclusions}

We have used a simple model to study the effect of the finite speed of light, and therefore time of flight of photons to an observer, and quantify how this affects the formation and persistence of spectral features in kilonova observations. We illustrate the bluer absorption dip of the P Cygni feature can form prior to the redder emission spike, as seen in the spectra of AT2017gfo, and discussed by \citet{sneppen2024emergencehourbyhourrprocessfeatures}. The emission component of this P Cygni feature persists for a long lifetime in comparison to the absorption component, with a significant contribution to the spectrum from scattered photons being visible from packets travelling for approximately 35~per~cent of the time since merger at each epoch. In this way, the reverberation effect due to the time of flight of the scattered and most significantly delayed photons will become stronger, and more significant, as time passes from the instant of the merger. As such, we reproduce a redward shift of the peak of spectral features with time, as is also identified in the X-shooter spectra of AT2017gfo (\citealt{Pian17, Smartt2017}).
We also find a redward drift of the scattering feature, although the emission becomes too faint by later times -- and predict this redshift due to the reverberation effect will be more prominent in any events observed with a more rapidly evolving ionisation balance. This reverberation effect contributes to, but cannot fully explain, the persistent nature of the spectral feature of AT2017gfo commonly attributed to \ion{Sr}{ii}. 

We also see the spectrum evolving on short time scales, with contributions from photons released from the photosphere within hours of each other having noticeably different characteristics. To this end, we agree with the conclusion of \citet{sneppen2024emergencehourbyhourrprocessfeatures}, namely that reverberation effects contribute to the prominent and persistent emission visible in the 1\,\micron\ feature of AT2017gfo. The nature of this feature is not attributable to one physical process and is instead a combination of many contributing factors; full analysis of these is beyond the scope of this work. 

Furthermore, we illustrate the importance of considering the effect of the travel time of light from an event to an observer when using time-independent codes to reproduce the thermal continuum. Whilst a simple shift in photospheric temperature can provide a reasonable approximation to accounting for this travel time effect when the photospheric temperature evolves slowly, it does not adequately correct the spectrum during periods of rapid evolution. For the 1.43 day observation of AT2017gfo, the temperature of the photosphere is cooling too rapidly for such a simple correction to be sufficient, and would require a more in-depth treatment. As such, if faster evolving events are observed in the future, we caution a more sophisticated temperature correction would be needed. 

Using a fully time-dependent code would be best practice when computational resources allow, but attention to potential differences between actual and observed photospheric temperatures in such events should always be employed, particularly during periods of especially rapid spectral evolution. Ultimately, including time-delay effects is increasingly important for rapidly expanding and quickly evolving transients, as they produce non-negligible effects that need to be modelled for fully accurate interpretations of the observed spectra.

\section*{Acknowledgements}

FM, SAS, RD and AS acknowledge funding by the European Union (ERC, HEAVYMETAL, 101071865). Views and opinions expressed are however those of the author(s) only and do not necessarily reflect those of the European Union or the European Research Council. Neither the European Union nor the granting authority can be held responsible for them. CEC is funded by the European Union’s Horizon Europe
research and innovation programme under the Marie Skłodowska-Curie grant agreement No. 101152610. LJS acknowledges support by the European Research Council (ERC) under the European Union's Horizon 2020 research and innovation program (ERC Advanced Grant KILONOVA No. 885281).

\section*{Data Availability}

The data underlying this article will be shared on reasonable request to the corresponding author.

\bibliographystyle{mnras}
\bibliography{example.bib} 

\appendix

\section{Trigonometry of the model} \label{app:geometry}

\begin{figure}
    \begin{center}
	\includegraphics[width=\columnwidth]{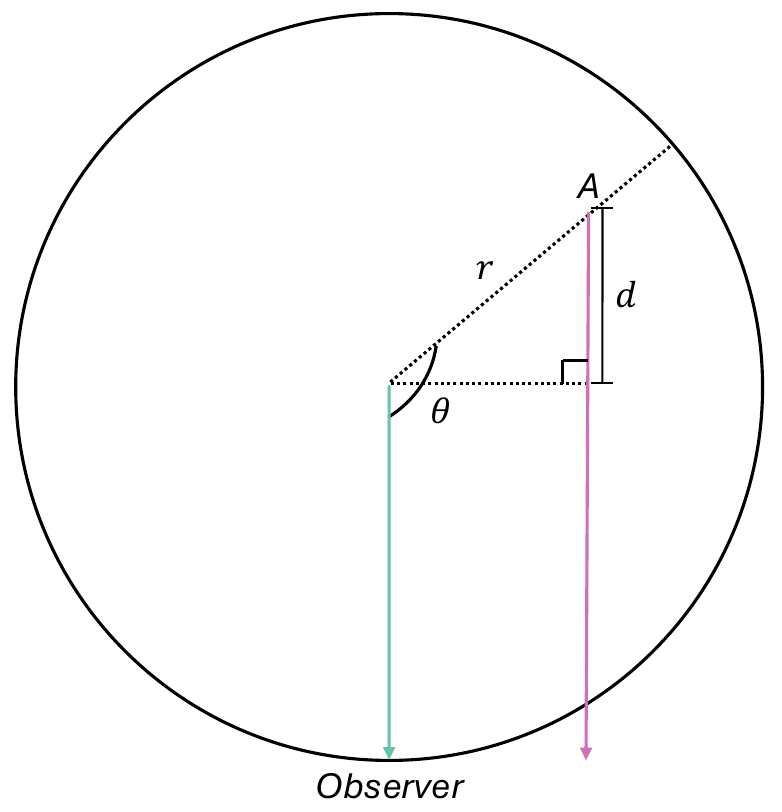}
    \end{center}
	\caption{The distances and angles used for calculating the distance a photon packet must travel to an observer located as shown. Two possible paths are shown: an unscattered photon travelling directly to the observer from the origin (teal) and a photon emerging into the observer's line of sight after being generated at point $A$, having travelled an additional distance $d$ (pink). The photosphere has been left off the diagram for simplicity.}
	\label{fig:trig}
\end{figure}

Each emergent photon packet will carry a frequency and trajectory, with two example photon packets shown in \autoref{fig:trig}. The teal arrow shows the reference case of a photon released from the origin which travels directly to the observer without scattering. The pink arrow shows a packet generated at point $A$, with a radial position $r$, which is emitted (or scattered) into the observer's line of sight. The trajectory is defined by a radial position and a direction cosine, and is used to calculate any additional distance a packet travels to the observer relative to the reference case. This direction cosine, $\mu_i$, as used in \autoref{eqn:new_nu}, is defined by the angle to the local radius vector. For the reference packet (teal), this angle is zero and for the pink packet is the corresponding angle to $\theta$ in \autoref{fig:trig}.

This simple trigonometry is used to calculate the relative additional distance, $d$, to the observer. By our definition, a photon packet generated at the origin has the delay time of zero (i.e. the packet depicted using a teal arrow in \autoref{fig:trig}).

The distance denoted $d$ for the pink packet in \autoref{fig:trig} is used to link $t_{\rm ph}$ and $t_{\rm det}$ through \autoref{eqn:time_relation} (i.e. the pink packet, as illustrated, has a positive delay time of $d/c$).

\bsp	
\label{lastpage}
\end{document}